\documentclass[12pt,a4paper]{article}

\usepackage{pslatex}	
\usepackage{amsfonts}
\usepackage{amsmath}
\usepackage{amsthm}
\usepackage{epsf}

\def\beq{\begin{equation}}
\def\eeq{\end{equation}}
\def\bea{\begin{eqnarray}}
\def\eea{\end{eqnarray}}
\let\nn=\nonumber
\def\beann{\begin{eqnarray*}}
\def\eeann{\end{eqnarray*}}

\let\a=\alpha \let\be=\beta \let\g=\gamma \let\de=\delta
\let\e=\varepsilon  \let\h=\eta 

\let\dh=\vartheta \let\k=\kappa \let\la=\lambda \let\m=\mu
\let\n=\nu  \let\p=\pi \let\r=\rho \let\s=\sigma
\let\om=\omega 
\let\ph=\varphi \let\Ph=\phi  \let\Ps=\Psi
  
 \let\G=\Gamma \let\D=\Delta

\let\qd=\quad  

\def\tst#1{{\textstyle #1}}

\theoremstyle{definition}

\def\2{\frac{1}{2}} \def\4{\frac{1}{4}}

\def\6{\partial}

\def\<{\langle} \def\>{\rangle}

\let\auf=\uparrow \let\ab=\downarrow

  \def\CO{{\cal O}}

\def\i{{\rm i}}

 \def\ch{{\rm ch}}

\def\sign{{\rm sign}} \def\End{{\rm End}}

\def\tr{{\rm tr}}

\begin{document}

\thispagestyle{empty}
\begin{center}
{\Large {\bf Universal correlations of one-dimensional electrons at
low density\\}}
\vspace{7mm}
{\large Frank G\"{o}hmann\footnote[2]{%
e-mail: Frank.Goehmann@uni-bayreuth.de}\\
\vspace{5mm}
Theoretische Physik I, Universit\"at Bayreuth, 95440 Bayreuth,
Germany\\}
\vspace{20mm}

{\large {\bf Abstract}}
\end{center}
\begin{list}{}{\addtolength{\rightmargin}{10mm}
               \addtolength{\topsep}{-5mm}}
\item
We summarize results on the asymptotics of the two-particle Green
functions of interacting electrons in one dimension. Below a critical
value of the chemical potential the Fermi surface vanishes, and the
system can no longer be described as a Luttinger liquid. Instead, the
non-relativistic Fermi gas with infinite point-like repulsion becomes
the universal model for the long-wavelength, low temperature physics
of the one-dimensional electrons. This model, which we call the
impenetrable electron gas, allows for a rigorous mathematical
treatment. In particular, a so-called determinant representation for
the two particle Green function could be derived. This determinant
representation is related to an integrable classical evolution
equation and to a Riemann-Hilbert problem, that enable the exact
calculation of the asymptotics of the two-particle Green functions.
\\[2ex]
\end{list}

\clearpage

\section{Correlation functions}
Correlation functions are (thermal) expectation values of products
of field operators. As a typical example we may consider the
two-particle Green function $G_{\a \be} (x, t) = \< \Ps_\a (x, t)
\Ps_\be^+ (0, 0) \>$. Here $\Ps_\a$ and $\Ps_\a^+$ are annihilation
and creation operators of electrons with spin $\a = \auf, \ab$. The
brackets denote the grand canonical ensemble average. The two-particle
Green function $G_{\a \be} (x, t)$ gives the probability to find an
electron at $(x, t)$ provided there was an electron at $(0, 0 )$. It
describes the propagation of an electron in a medium of other
electrons, which, in general, is characterized by a temperature $T$,
a chemical potential $\m$ and a magnetic field~$B$.

Within the frame work of the linear response theory correlation
functions provide the link between the microscopic and the macroscopic
properties of matter. Correlation functions are measurable in
experiments. The two-particle Green function, for instance, measures
the optical absorption.

Only a few mathematically exact results on correlation functions of
interacting systems are known. In fact, the example discussed in this
contribution is the first example of a direct Bethe ansatz calculation
of the asymptotics of correlations of interacting electrons. The
calculation was performed for the so-called impenetrable electron gas
model, which is the infinite repulsive coupling limit of Yang's model
of electrons with point-like pair interaction \cite{Yang67,Gaudin67}.
As we shall argue below, our results are, in spite of being obtained
for a specific model, applicable to a whole class of models of interacting
electrons in one dimension, and are, in this sense, universal.

In the first part of this text we shall explain our ideas about the
universality of the long-wavelength, low-temperature asymptotics
of electronic correlations in the gas phase \cite{GoKo99a}. We
shall start with the paradigmatic Hubbard model, and then argue that
certain modifications of the interaction do not change the 
long-wavelength, low-temperature physics of the model. In the appropriate
scaling limit all modified Hamiltonians lead to the same effective model.
The second part of this text is devoted to a summary of our calculation
of the asymptotics of the two-particle Green functions of the impenetrable
electron gas model \cite{GIKP98,GIK98}.

\section{The Hubbard model in the gas phase}
The density $D$ of non-interacting, one-dimensional spin-$\2$ Fermions
on a lattice is given by the integral over the Fermi weight,
\begin{equation}
     D = \frac{2}{\p} \int_0^\p \! dp \,
         \frac{1}{e^{(\e(p) - \m)/T} + 1}\ .
\end{equation}
Here $\e(p)$ is the dispersion of the Fermions, $T$ denotes the temperature
and $\m$ the chemical potential. Let us assume $\e(p)$ to be monotonically
increasing and bounded from below. If the chemical potential is smaller
than a critical value, $\m < \m_c = \min_{p > 0} \e(p)$, then $D$ vanishes
in the zero temperature limit $T \rightarrow 0+$. For $\m > \m_c$, on
the other hand, the density $D$ approaches a finite positive value as
$T \rightarrow 0+$. This means that the system undergoes a phase
transition at $T = 0$ as a function of the chemical potential. The
critical point is at $\m = \m_c$. Assuming that $\e(p) = \m_c + p^2 +
{\cal O}(p^4)$, we obtain
\begin{equation} \label{densefree}
     D = \frac{2}{\p} \sqrt{\m - \m_c}
\end{equation}
for $\m_c < \m < \m_c + \de$, $\de \ll 1$.

As we shall see below by considering a representative example this
scenario remains unchanged if the Fermions interact. For interacting
one-dimensional Fermions the phase with $\m > \m_c$, for which the
density at $T = 0$ is finite, is called the Luttinger liquid phase
\cite{Haldane80,Haldane81,Voit94}. This phase is quite familiar to many
physicists. A Luttinger liquid may be understood as a one-dimensional
metal. The correlations in the Luttinger liquid are dominated by
fluctuations around the Fermi surface. Their power law decay at zero
temperature is described by conformal field theory \cite{BPZ84}.
For small finite temperature conformal field theory predicts
exponentially decaying correlations. One has to employ a conformal
mapping from the complex plane to a strip of finite width. As a result
the rate of exponential decay is defined by conformal dimensions.

The phase with $\m < \m_c$ so far attracted less attention. This phase
is trivial at $T = 0$, since the density $D$ vanishes for $T = 0$. The
density becomes positive for positive temperature, and is typically
exponentially small as long as the temperature remains small. The ideal
gas law holds. This suggested the name `gas phase' to us. The gas phase
may be interpreted as a one-dimensional semi conductor or insulator.
Correlations in the gas phase behave essentially different compared to
those in the Luttinger liquid phase. This is the subject of this text.

In order to get a better understanding of the gas phase of interacting
systems let us consider the Hubbard model as an example,
\begin{equation}
     H_H = - \sum_{j=1}^L (c_{j+1, \sigma}^+ c_{j, \sigma} +
			 c_{j, \sigma}^+ c_{j+1, \sigma})
         + U \sum_{j=1}^L n_{j \uparrow} n_{j \downarrow}
	    - \mu \sum_{j=1}^L (n_{j \uparrow} + n_{j \downarrow})\ .
\end{equation}
Here the canonical Fermi operators $c_{j, \sigma}^+$, $c_{j, \sigma}$
are creation and annihilation operators of electrons at site $j$
of a one-dimensional, periodically closed chain of $L$ lattice sites, and
$n_{j, \uparrow}$ and $n_{j, \downarrow}$ are the corresponding particle
number operators. $U$ is the strength of the repulsive interaction.

The eigenvalue problem of the Hubbard Hamiltonian can be solved
\cite{LiWu68} by means of the nested Bethe ansatz. This allows us to
test our ideas about the gas phase quantitatively. The energy levels for
the $N$ electron system are
\begin{equation} \label{energy}
     E = 2 \sum_{j=1}^N (1 - \cos k_j) - (\mu + 2) N\ ,
\end{equation}
where the charge momenta $k_j$ are solutions of the Lieb-Wu equations
\cite{LiWu68}. Clearly the first term on the right hand side of
(\ref{energy}) is non-negative. Hence, if $\mu < \mu_c = - 2$, the
energies of all eigenstates become non-negative, and the absolute ground
state is the empty lattice. For $\mu > - 2$, on the other hand, the
energy can be lowered by filling states with small $k$'s. Since
$k_{j+1} - k_j \sim 1/L$, this leads to a finite density of electrons
in the ground state as $L \rightarrow \infty$. We conclude that the
Hubbard model is in the gas phase for $\mu < - 2$ and in the Luttinger
liquid phase else. Note that the asymptotics of correlation functions of the
Hubbard model in the Luttinger liquid phase was obtained in
\cite{FrKo90,FrKo91}.

Another way of understanding the transition from the gas phase to the
Luttinger liquid phase is by looking at the integral equations that
describe the Hubbard model in the thermodynamic limit (see e.g.\
\cite{DEGKKK00}). At zero temperature and zero magnetic field the dressed
energy $\k(k)$ of the elementary charge excitations is determined by the
integral equation
\begin{equation} \label{kappanull}
     \k(k) = - 2 \cos(k) - \m  + \int_{-Q}^Q \! dk' \cos(k')
               R\big(\sin(k') - \sin(k)\big) \k(k')\ ,
\end{equation}
where the limits of integration depend on the chemical potential
through the condition
\begin{equation} \label{fermiimp}
     \k(Q) = 0\ .
\end{equation}
The integral kernel $R$ is given by
\begin{equation} \label{hubkernel}
     R(x) = \int_{- \infty}^\infty \! \frac{d \om}{2 \p}
               \frac{e^{\i \om x}}{1 + e^{U |\om|/2}}\ .
\end{equation}
Similarly, the density $\r(k)$ of elementary charge excitations is obtained
from the integral equation
\begin{equation} \label{rhonull}
     \r(k) = \frac{1}{2 \p}  + \int_{-Q}^Q \! dk' \cos(k)
               R\big(\sin(k') - \sin(k)\big) \r(k')\ .
\end{equation}
$\r(k)$ determines the density of the electrons at zero temperature,
\begin{equation} \label{densenull}
     D = \int_{-Q}^Q \! dk \, \r(k)\ .
\end{equation}
Since $\r(k)$ is positive, $D$ is equal to zero if and only if $Q = 0$.
Hence, it follows from (\ref{kappanull}) and (\ref{fermiimp}) that
$\m_c = -2$. Moreover, the equations (\ref{kappanull})-(\ref{densenull})
are easily solved for small Q. From the solution we obtain
\begin{equation} \label{denseint}
     D = \frac{1}{\p} \sqrt{\m - \m_c}
\end{equation}
for $\m_c < \m < \m_c + \de$, $\de \ll 1$.

Thus the qualitative picture is the same for free and for interacting
electrons. Comparing (\ref{densefree}) and (\ref{denseint}), however, we
see that the results for the electron density close to the critical point
$\m = \m_c$ differ by a factor of two. This difference may be interpreted
as a signature of a phenomenon called spin-charge separation (see, for
instance, \cite{Voit94}): The elementary charge excitations of the Hubbard
model at finite positive $U$ are spinless, hence the density is smaller by
a factor of two.

To complete our understanding of the gas phase of the one-dimensional
Hubbard model let us consider the low temperature thermodynamics of the gas
phase. The thermodynamics of the Hubbard model was first considered by
Takahashi \cite{Takahashi72}. The limiting case we are interested in,
however, was not studied in Takahashi's paper. Takahashi expressed the Gibbs
free energy $\omega = - P$ ($P$ pressure) in terms of the dressed energies
$\kappa (k)$, $\varepsilon_n (\Lambda)$, $\varepsilon_n' (\Lambda)$, of
elementary excitations at finite temperature. $\kappa (k)$ is the
dressed energy of particle (or hole) excitations, $\varepsilon_n
(\Lambda)$ describes spin excitations and $\varepsilon_n' (\Lambda)$
so-called $k$-$\Lambda$ strings \cite{DEGKKK00}. All $k$-$\Lambda$ strings
are gapped \cite{Takahashi72}. They do not contribute to the low-temperature
thermodynamic properties of the Hubbard model \cite{Takahashi74} and
drop out of the equation for the pressure, which simplifies to
\begin{equation}
     P = \, \frac{T}{2 \pi} \int_{- \p}^\p dk \,
	 \ln \left( 1 + e^{- \frac{\kappa (k)}{T}} \right)\ .
\end{equation}
Similarly, the integral equations for the dressed energies at low
temperature become
\begin{align}
     \kappa (k) = & - \mu - 2 \cos k 
		     - \, T \sum_{n=1}^\infty
		     \left( [n] \ln \left( 1 +
			e^{- \frac{\varepsilon_n}{T}} \right) \right)
			(\sin k)\ , \label{kappa} \\
     \ln \left( 1 + e^{\frac{\varepsilon_n (\Lambda)}{T}} \right) = &
	   - \int_{- \pi}^\pi dk \, \cos k \, a_n (\Lambda - \sin k)
	     \ln \left( 1 + e^{- \frac{\kappa (k)}{T}} \right) \nonumber
	\\ & + \sum_{m=1}^\infty \left( A_{nm}
	     \ln \left( 1 + e^{- \frac{\varepsilon_n}{T}} \right)
	     \right) (\Lambda)\ , \label{epsn}
\end{align}
where $n = 1, 2, 3, \dots$ in equation (\ref{epsn}), and
\begin{equation}
     a_n (\Lambda) = \frac{nU/4 \pi}{(nU/4)^2 + \Lambda^2}\ .
\end{equation}
$[n]$ and $A_{nm}$ are integral operators defined by
\begin{align}
     ([0]f) (\Lambda) = & f(\Lambda)\ , \\
     ([n]f) (\Lambda) = & \int_{- \infty}^\infty d \Lambda' \,
	a_n (\Lambda - \Lambda') f(\Lambda')\ , \ n = 1, 2, \dots \\
     A_{nm} = & \sum_{j=1}^{{\rm min} \{n, m\}}
		 \left( \big[|n - m| + 2(j - 1) \big] +
		 \big[|n - m| + 2j \big] \right).
\end{align}
The gas phase is characterized by the absence of a Fermi surface for
$\kappa (k)$. Thus $\kappa (k)$ is positive in the zero temperature
limit, and the first term on the right hand side of (\ref{epsn}) becomes
exponentially small in $T$. Dropping this term, the equations
(\ref{epsn}) decouple from (\ref{kappa}). Since the equations become
independent of $\Lambda$, it is not hard to solve them. The solution,
$\exp\{ \varepsilon_n (\Lambda)/ T\} = n(n + 2)$, is the same as in the
infinite coupling limit $U \rightarrow \infty$ (cf.\ e.g.\
\cite{Takahashi71b}). Inserting this solution into (\ref{kappa}) we obtain
\begin{equation} \label{kappagas}
     \kappa (k) = - \mu - 2 \cos k - T \ln 2\ .
\end{equation}
Our initial assumption, that $\lim_{T \rightarrow 0} \kappa (k) > 0$ holds
for all $k$, is self-consistent, if $\mu + 2 < 0$, which is precisely the
condition for being in the gas phase stated above. With (\ref{kappagas})
the low temperature expression for the pressure becomes
\begin{equation} \label{gibbshu}
     P = \, \frac{T}{2 \pi} \int_{- \p}^\p dk \,
      	 \ln \left( 1 + 2 e^{\frac{\mu + 2 \cos k}{T}} \right)
            \approx \, \sqrt{\frac{T}{\pi}} \, e^{\frac{\mu + 2}{T}}\ ,
\end{equation}
and we see that the density $D = \partial P/\partial \mu$ and the pressure
$P$ are related by the ideal gas law,
\begin{equation} \label{ideal}
     P = TD\ .
\end{equation}

There are two important lessons to learn from our simple calculation.
First, the low temperature limit in the gas phase works the same way
as the strong coupling limit at finite temperatures. Second, the low
temperature Gibbs free energy $\omega = - P$ in the gas phase shows no
signature of the discreteness of the lattice. It is the same as for the
impenetrable electron gas (see below), which is a continuum model. This
agrees well with our intuitive understanding of the gas phase at low
temperature: (i) The mean free path ($= 1/D$) of the electrons is large
compared to the lattice spacing (which we set equal to unity so far).
(ii) Their kinetic energy is of the order $T$. Hence, the effective
repulsion is large for $T \ll U$. (iii) The ideal gas law holds at low
temperature.

\section{Scaling}
The above arguments show that only electrons with small momenta,
corresponding to long wavelengths contribute to the low-temperature
properties of the Hubbard model in the gas phase. Thus the Hubbard model
in the gas phase at low temperature is effectively described by its
continuum limit. In order to perform the continuum limit we have to
introduce the lattice spacing $\D$ and coordinates $x = \Delta n$
connected with the $n$th lattice site. The total length of the system
is $\ell = \D L$. The continuum limit is the limit $\Delta \rightarrow 0$
for fixed $\ell$. In this limit we obtain canonical field operators
$\Psi_\sigma (x)$ for electrons of spin $\sigma$ as
\begin{equation} \label{psi}
     \Psi_\sigma (x) = \lim_{\Delta \rightarrow 0} \,
		       c_{n, \sigma}/\sqrt{\Delta}\ .
\end{equation}
Let us perform the rescaling
\begin{equation}
     T_H = \Delta^2 T\ ,\ \mu_H + 2 = \Delta^2 \mu\ ,\ k_H = \Delta k\
     ,\ t_H = t/\Delta^2\ ,\ B_H = \Delta^2 B\ , \label{scale}
\end{equation}
where $k$ denotes the momentum, $t$ the time and $B$ the magnetic field,
which we shall incorporate below. The index `$H$' refers to the Hubbard
model. Then, in the limit $\Delta \rightarrow 0$, we find
\begin{equation}
     H_H/T_H = H/T\ .
\end{equation}
Here $H$ is the Hamiltonian for continuous electrons with delta
interaction,
\begin{multline}
     H = \int_{- \ell/2}^{\ell/2} dx \Big\{
            (\partial_x \Psi_\alpha^+ (x)) \partial_x \Psi_\alpha (x)
	    + \frac{U}{\Delta} \Psi_\uparrow^+ (x) \Psi_\downarrow^+ (x)
	      \Psi_\downarrow (x) \Psi_\uparrow (x) \\
	    - \, \mu \, \Psi_\alpha^+ (x) \Psi_\alpha (x)\Big\} \quad.
	    \label{gasham}
\end{multline}
Note that the coupling $c_1 = U/\Delta$ of the continuum model goes to
infinity! This is a peculiarity of the one-dimensional system. The
effective interaction in the low density phase becomes large. Similar
scaling arguments lead to an effective coupling $c_2 = U$ in two
dimensions and to $c_3 = \Delta U$ in three dimensions, i.e.\ unlike
one-dimensional electrons three-dimensional electrons in the gas phase
are free.

\section{Universality}
What happens to more general Hamiltonians in the continuum limit?
Let us consider Hamiltonians of the form $H_G = H_H + V$, where
$H_H$ is the Hubbard Hamiltonian and $V$ contains additional short
range interactions. We shall assume that $V$ is a sum of local
terms $V_j$ which preserve the particle number. Then $V_j$ contains as
many creation as annihilation operators, and the number of field
operators in $V_j$ is even. We shall further assume that $V_j$ is
hermitian and space parity invariant.

According to equation (\ref{psi}) every field $c_{j, \sigma}$ on the
lattice contributes a factor of $\Delta^{1/2}$ in the continuum limit.
One factor of $\Delta$ is absorbed by the volume element $dx = \Delta$,
when turning from summation to integration. Thus, if $V_j$ contains 8
or more field operators, then $V \sim \Delta^3$ and $V/T_H$ vanishes.
If $V_j$ contains 6 field operators, then at least two of the creation
operators and two of the annihilation operators must belong to different
lattice sites, since otherwise $V_j = 0$. A typical term is, for instance,
$V_j = c_{j, \uparrow}^+ c_{j, \downarrow}^+ c_{j + 1, \uparrow}^+
c_{j + 1, \uparrow} c_{j, \downarrow} c_{j, \uparrow}$. In the continuum
limit we have $c_{j + 1, \uparrow} = \Delta^{1/2} \Psi_\uparrow (x)
+ \Delta^{3/2} \partial_x \Psi_\uparrow (x) + O (\Delta^{5/2})$.
Hence, the leading term vanishes due to the Pauli principle. The next
to leading term acquires an additional pow\-er of $\Delta$. We conclude
that $V \sim \Delta^3$ and thus $V/T_H \rightarrow 0$.

If $V_j$ contains 4 fields, then
\begin{equation} \label{4fields}
     V \sim \Delta^2 \Psi_\uparrow^+ (x) \Psi_\downarrow^+ (x)
	\Psi_\downarrow (x) \Psi_\uparrow (x) + O (\Delta^4)\ .
\end{equation}
Here the first term on the right hand side is the density-density
interaction of the electron gas. In order to arrive at the impenetrable
electron gas model the coefficient in front of this term has to be
positive. Note that there are no terms of the order of $\Delta^3$ on
the right hand side of (\ref{4fields}) and thus no other terms than
the first one in the continuum limit. Terms of the order of $\Delta^3$
would contain precisely one spatial derivative. They are ruled out,
since they would break space parity.

Considering the case, when $V_j$ contains 2 fields, we find, except
for the kinetic energy and the chemical potential term, terms which
correspond to a coupling to an external magnetic field $B_H$. For these
terms to be finite in the continuum limit we have to rescale the
magnetic field as $B_H = \Delta^2 B$ (cf.\ equation (\ref{scale})).

Our considerations show that the impenetrable electron gas model with
magnetic field,
\begin{equation} \label{hb}
     H_B = H + B \int_{- \ell/2}^{\ell/2} dx \;
	   \Psi_\alpha^+ (x) \sigma_{\alpha \beta}^z \Psi_\beta (x)\ ,
\end{equation}
is indeed the universal model (for small $T$) for the gas phase of
one-dimensional lattice electrons with repulsive short-range
interaction.

\section{Impenetrable electrons}
The impenetrable electron gas is the infinite coupling limit of the
electron gas with repulsive delta interaction ($\Delta \rightarrow 0$
in (\ref{hb})), which was the first model solved by nested Bethe ansatz
\cite{Yang67,Gaudin67}. The pressure of the system as a function of $T$,
$\mu$ and $B$ is known explicitly \cite{Takahashi71b},
\begin{equation} \label{gibbsif}
     P = \, \frac{T}{2 \pi} \int_{- \infty}^\infty dk \,
	 \ln \left( 1 +  e^{\frac{\mu + B - k^2}{T}}
		             +  e^{\frac{\mu - B - k^2}{T}} \right)\ ,
\end{equation}
and may serve as thermodynamic potential. The expression (\ref{gibbsif})
is formally the same as for a gas of free spinless Fermions with
effective (temperature dependent) chemical potential $\mu_{eff} = \mu +
T \ln(2 \cosh B/T)$. Hence the Fermi surface vanishes for
$\lim_{T \rightarrow 0} \mu_{eff} = \mu + |B| < 0$. The finite
temperature correlation functions of the impenetrable electron gas
depend crucially on the sign of $\mu_{eff}$. This allows us to define
the gas phase at finite temperature by the condition $\mu_{eff} < 0$,
which is also sufficient for deriving the ideal gas law (\ref{ideal})
from the low temperature limit of (\ref{gibbsif}). Note that for zero
magnetic field and small temperature equation (\ref{gibbsif}) coincides
with the right hand side of (\ref{gibbshu}).

The time and temperature dependent (two-point) Green functions are
defined as
\begin{eqnarray} \label{gp1}
     G_{\uparrow \uparrow}^+ (x,t) & = &
		 \frac{{\rm tr} \left( e^{- H_B/T} \,
        \Psi_\uparrow (x,t) \Psi_\uparrow^+ (0,0) \right)}
	{{\rm tr} \left( e^{- H_B/T} \right)}\ , \\ \label{gm1}
     G_{\uparrow \uparrow}^- (x,t) & = &
		 \frac{{\rm tr} \left( e^{- H_B/T} \,
        \Psi_\uparrow^+ (x,t) \Psi_\uparrow (0,0) \right)}
	{{\rm tr} \left( e^{- H_B/T} \right)}\ .
\end{eqnarray}
For the impenetrable electron gas these correlation functions were
represented as determinants of Fredholm integral operators in
\cite{IzPr97,IzPr98}. The determinant representation provides a powerful tool
to study their properties analytically.

In \cite{GIKP98,GIK98} the determinant representation was used to
derive a nonlinear partial differential equation for two classical
auxiliary fields, which determine the correlation functions. This
partial differential equation is closely related to the Heisenberg
equation of the quantum Hamiltonian (\ref{gasham}). It is called the
separated nonlinear Schr\"odinger equation. Together with a corresponding
Riemann-Hilbert problem it determines the large-time, long-distance
asymptotics of the correlators (\ref{gp1}), (\ref{gm1}) (for details see
the following sections). In \cite{GIKP98,GIK98} the asymptotics $x, t
\rightarrow \infty$ was calculated for fixed ratio $k_0 = x/2t$. The
crucial parameter for the asymptotics is the average number of particles
$xD$ in the interval $[0,x]$. If $x$ is large but $xD \ll 1$
(i.e.\ $T$ small), an electron propagates freely from $0$ to $x$, and the
correlation functions (\ref{gp1}), (\ref{gm1}) are those of free Fermions,
\begin{eqnarray} \label{gpf}
     G_f^+ (x,t) & = & \frac{e^{- \frac{{\rm i} \pi}{4}}}{2 \sqrt{\pi}}
		       \, t^{- \frac{1}{2}} e^{{\rm i} t (\mu - B)}
			  e^{\frac{{\rm i} x^2}{4t}}\ , \\ \label{gmf}
     G_f^- (x,t) & = & \frac{e^{\frac{{\rm i} \pi}{4}}}{2 \sqrt{\pi}}
	               \, e^{\frac{(\mu - B - k_0^2)}{T}}
		       \, t^{- \frac{1}{2}} e^{- {\rm i} t (\mu - B)}
			  e^{- \frac{{\rm i} x^2}{4t}}\ .
\end{eqnarray}
The true asymptotic region is characterized by a large number $xD$ of
particles in the interval $[0,x]$, specifically, $xD \gg z_c^{- 1}$,
where $z_c = (T^{3/4} e^{- k_0^2/2T})/(2 \pi^{1/4} k_0^{3/2})$. If the
latter condition is satisfied, the correlation functions decay due to
multiple scattering. The cases $B > 0$ and $B \le 0$ have to be treated
separately. For $B > 0$ a critical line, $x = 4t \sqrt{B}$, separates
the $x$-$t$ plane into a time and a space like regime. The asymptotics
(for small $T$) in these respective regimes are:\\[1ex]
{\em Time like regime ($x < 4t \sqrt{B}$):}
\begin{equation} \label{gpmas2}
     G_{\uparrow \uparrow}^\pm (x,t) = G_f^\pm (x,t) \,
	\frac{t^{\mp {\rm i} \nu(z_c)} e^{- x D_\downarrow}}
	     {\sqrt{4 \pi z_c x D_\downarrow}}\ ,
\end{equation}
where
\begin{equation} \label{nuzp}
     \nu(z_c) = - \; \frac{2D_\downarrow k_0^{3/2} e^{- k_0^2/2T}}
                              {\pi^{1/4} T^{5/4}}\qd, \qd
     D_\downarrow = \frac{1}{2} \sqrt{\frac{T}{\pi}}
				    e^{(\mu + B)/T}\ .
\end{equation}
$D_\downarrow = \partial P /\partial (\mu + B)$ is the low temperature
expression for the density of down-spin electrons.\\[1ex]
{\em Space like regime ($x > 4t \sqrt{B}$):}
\begin{equation} \label{gpmas}
     G_{\uparrow \uparrow}^\pm (x,t) = G_f^\pm (x,t) \,
	t^{\mp {\rm i} \nu(\gamma^{-1})} e^{- x D_\downarrow}\ ,
\end{equation}
where
\begin{equation} \label{nugamma}
     \nu(\gamma^{-1}) = - \; \frac{e^{(3B + \mu - k_0^2)/T}}{2\pi}\ .
\end{equation}
For $B \le 0$ there is no distinction between time and space like
regimes. The asymptotics is given by (\ref{gpmas}).

It is fair to mention here that the calculation of the asymptotics
(\ref{gpmas2}), (\ref{gpmas}) is rather lengthy. Equations (\ref{gpmas2})
and (\ref{gpmas}) are asymptotic expansions in $t$ consisting of an
exponential factor, a power law factor and a constant factor. Note that
the method employed in \cite{GIK98} allows for a systematic calculation
of the next, subleading orders.

The leading exponential factor in (\ref{gpmas2}) and
(\ref{gpmas}), has a clear physical interpretation: Because
of the specific form of the infinite repulsion in (\ref{gasham}),
up-spin electrons are only scattered by down-spin electrons. This is
reflected in the fact that the correlation length is $1/D_\downarrow$.
The expression $1/D_\downarrow$ may be interpreted as the mean free path
of the up-spin electrons. Thus the correlation length for up-spin
electrons is equal to their mean free path. The exponential decay of the
two-particle Green functions means that, due to the strong interaction,
an up-spin electron is confined by the cloud of surrounding down-spin
electrons. Thus we are facing an interesting situation: Although at
small distances the electrons look like free Fermions, they are
confined on a macroscopic scale set by the mean free path $1/D_\downarrow$.

\section{Outline of the derivation}
The derivation of the above results on the asymptotics of the two-particle
Green functions at low temperature is based on the fact that the
impenetrable electron gas model is exactly solvable by Bethe ansatz.
The Bethe ansatz eigenfunctions \cite{Yang67,Gaudin67} and the
thermodynamics of the model \cite{Takahashi71b} are known since long.
But only recently a determinant representation for the two-particle
Green functions was derived by Izergin and Pronko \cite{IzPr97,IzPr98}.
Their derivation includes the following steps:
\begin{enumerate}
\item
A change of basis for the spin part of the Bethe ansatz wave function
from inhomogeneous XXX to XX spin chain eigenfunctions, which is possible
at infinite repulsion.
\item
The calculation of form factors in the finite volume.
\item
A summation of the form factors.
\item
The thermodynamic limit.
\end{enumerate}
The details of the calculation can be found in the article
\cite{IzPr98}.

The asymptotic analysis of the correlation functions was performed
in \cite{GIKP98,GIK98}. Starting point was the determinant
representation of Izergin and Pronko (section~\ref{sec:detrep}),
which is valid for all $x$ and $t$. The non-trivial ingredients of
the determinant representation are certain auxiliary functions $b_{++}$
and $B_{--}$ and the Fredholm determinant $\det(\hat I +\hat V)$ of an
integral operator $\hat V$. A direct, yet lengthy calculation shows that
$b_{++}$ and $B_{--}$ satisfy the separated non-linear Schr\"odinger
equation (section~\ref{sec:nls}), which is a well-known integrable
partial differential equation. The logarithm of the Fredholm determinant
plays the role of its tau-function (section~\ref{sec:taufun}). Moreover,
a Riemann-Hilbert problem that fixes $b_{++}$ and $B_{--}$ as solutions
of the separated non-linear Schr\"odinger equation can be derived
from the determinant representation (section~\ref{sec:rhp}). The
Riemann-Hilbert problem is the appropriate starting point for the
asymptotic analysis of the correlation functions $G_{\auf \auf}^\pm$
(section~\ref{sec:assanal}).

Luckily, the differential equation and the Riemann-Hilbert problem
turn out to be of the same form as in case of the impenetrable
(spinless) Bose gas \cite{IIKS90,KBIBo}. Therefore a theorem obtained
in the asymptotic analysis of the impenetrable Bose gas \cite{IIKV92}
could be applied to the impenetrable electrons as well. In contrast to
the bosonic case, there is, however, an additional external integration
in the determinant representation of the impenetrable electron gas.
This integration can be carried out in the low temperature limit, by
the method of steepest descent (section~\ref{sec:steepest}).

\section{Determinant representation} \label{sec:detrep}
Let us recall the determinant representation for the correlation
functions $G_{\auf \auf}^\pm (x,t)$, which was derived in
\cite{IzPr97,IzPr98}. We shall basically follow the account of
\cite{GIKP98}. Yet, it turns out to be useful for further
calculations to rescale the variables and the correlation functions.
The rescaling
\begin{align} \label{xtr}
     & x_r = - \sqrt{T} x /2 \qd, \qd
       t_r = T t /2\ , \\
     & g^\pm = G_{\auf \auf}^\pm /\sqrt{T}\ , \\
     & \be = \m_{eff}/T \qd, \qd h = B/T
\end{align}
removes the explicit temperature dependence from all expressions.
Furthermore, it will allow us to make close contact with results which
were obtained for the impenetrable Bose gas \cite{IIKS90,IIKV92,KBIBo}.
The index `$r$' in (\ref{xtr}) stands for `rescaled'. For the sake of
simplicity we shall suppress this index in the following sections.
We shall come back to physical space and time variables only in the
last section, where we consider the low temperature limit.

The rescaled correlation functions $g^+$ and $g^-$ in the rescaled
variables can be expressed as \cite{GIKP98,GIK98},
\begin{align} \label{gp2}
     g^+ (x,t) = &
        \frac{- e^{2 \i t (\be - h - \ln(2\ch (h)))}}{2 \p}
        \int_{- \p}^\p \! d\h \, \frac{F(\g,\h)}{1 - \cos(\h)} \,
	b_{++} \det\left( \hat I + \hat V \right)\ , \\
     \label{gm2}
     g^- (x,t) = &
	\frac{e^{- 2 \i t (\be - h - \ln(2\ch (h)))}}{4 \p \g}
        \int_{- \p}^\p \! d\h \, F(\g,\h) B_{--}
	\det\left( \hat I + \hat V \right)\ .
\end{align}
Here $\g$ and $F(\g,\h)$ are elementary functions,
\begin{align}
     \g = & 1 + e^{2h}\ , \\
     F(\g,\h) = & 1 + \frac{e^{\i \h}}{\g - e^{\i \h}} +
                    \frac{e^{- \i \h}}{\g - e^{- \i \h}}\ .
\end{align}
$\det(\hat I + \hat V)$ is the Fredholm determinant of the integral
operator $\hat I + \hat V$, where $\hat I$ is the identity operator,
and $\hat V$ is defined by its kernel $V(\la,\m)$. $\la$ and $\m$ are
complex variables, and the path of integration is the real axis. In
order to define $V(\la,\m)$ we have to introduce certain auxiliary
functions. Let us define
\begin{align}
     \tau (\la) = & \i(\la^2 t + \la x)\ , \\
     \dh(\la) = & \frac{1}{1 + e^{\la^2 - \be}}\ , \\
         \label{eoflambda} \displaybreak[0]
     E(\la) = & \mbox{p.v.} \int_{- \infty}^\infty \! d\m \,
           \frac{e^{- 2 \tau(\m)}}{\p (\m - \la)}\ ,
	   \\ \label{em} \displaybreak[0]
     e_-(\la) = &
        \sqrt{\frac{\dh(\la)}{\p}} e^{\tau(\la)}\ , \\ \label{ep}
     e_+(\la) = &
        \frac{1}{2} \sqrt{\frac{\dh(\la)}{\p}} e^{-\tau(\la)}
           \left\{ (1 - \cos(\h)) e^{2 \tau(\la)} E(\la) + \sin(\h)
	      \right\}\ .
\end{align}
Note that $\dh(\la)$ is the Fermi weight. $V(\la,\m)$ can be expressed
in terms of $e_+$ and~$e_-$,
\begin{equation}
     V(\la,\m) = \frac{e_+(\la) e_-(\m) - e_+(\m) e_-(\la)}{\la - \m}\ .
\end{equation}

Denote the resolvent of $\hat V$ by $\hat R$,
\begin{equation}
     \left( \hat I + \hat V \right) \left( \hat I - \hat R \right) =
     \left( \hat I - \hat R \right) \left( \hat I + \hat V \right)
        = \hat I\ .
\end{equation}
Then $\hat R$ is an integral operator with symmetric kernel \cite{IIKS90},
\begin{equation} \label{kerr}
     R(\la, \m) = \frac{f_+(\la) f_-(\m) - f_+(\m) f_-(\la)}{\la - \m}\ ,
\end{equation}
which is of the same form as $V(\la,\m)$. The functions $f_\pm$ are
obtained as the solutions of the integral equations
\begin{equation} \label{fies}
     f_\pm(\la) + \int_{-\infty}^\infty \! d\m \, V(\la, \m) f_\pm(\m)
        = e_\pm (\la)\ .
\end{equation}

We may now define the potentials
\begin{equation} \label{defpot}
     B_{ab} = \int_{- \infty}^\infty \! d\la \, e_a(\la) f_b(\la) \qd,
              \qd
     C_{ab} = \int_{- \infty}^\infty \! d\la \, \la e_a(\la) f_b(\la)
\end{equation}
for $a, b = \pm$. $B_{--}$ enters the definition of $g^- (x,t)$,
equation (\ref{gm2}). $b_{++}$ in (\ref{gp2}) is defined as
\begin{equation}
     b_{++} = B_{++} - G(x,t)\ ,
\end{equation}
where
\begin{equation}
     G(x,t) = \frac{(1 - \cos(\h)) e^{- \i \p /4}}{2 \sqrt{2 \p t}}
	          e^{\i x^2/2t}\ .
\end{equation}
The remaining potentials $B_{ab}$ and $C_{ab}$ will be needed later.

It is instructive to compare the determinant representation (\ref{gp2})
for the correlation function $g^+ (x,t)$ with the corresponding
expression for impenetrable Bosons (cf e.g.\ page 345 of \cite{KBIBo}).
The main formal differences are the occurrence of the $\h$-integral in
(\ref{gp2}) and the occurrence of $\h$ in the definition of $e_+$. As
can be seen from the derivation of (\ref{gp2}) in \cite{IzPr98}, the
$\h$-integration is related to the spin degrees of freedom. For
$\h = \pm \p$ the expression $- \2 e^{2 \i \be t} b_{++}
\det(\hat I + \hat V)$ agrees with the field-field correlator for
impenetrable Bosons (recall, however, the different physical meaning
of~$\be$).

\section{Differential equations} \label{sec:nls}
As in case of impenetrable Bosons \cite{IIKS90,KBIBo} it is possible
to derive a set of integrable nonlinear partial differential equations
for the potentials $b_{++}$ and $B_{--}$ and to express the logarithmic
derivatives of the Fredholm determinant $\det(\hat I + \hat V)$ in
terms of the potentials $B_{ab}$ and $C_{ab}$.

The functions $f_\pm$ satisfy linear differential equations with respect
to the variables $x$, $t$, and $\be$,
\begin{equation} \label{lindiff}
     \hat L {f_+ \choose f_-} = \hat M {f_+ \choose f_-} =
        \hat N {f_+ \choose f_-} = 0\ ,
\end{equation}
The Lax operators $\hat L$, $\hat M$ and $\hat N$ are given as
\begin{align}
     \hat L = & \6_x + \i \la \s^z - 2 \i Q\ , \\
     \hat M = & \6_t + \i \la^2 \s^z - 2 \i \la Q + \6_x U\ , \\
     \hat N = & 2 \la \6_\be + \6_\la + 2 \i t \la \s^z + \i x \s^z
                  - 4 \i t Q - 2 \6_\be U\ ,
\end{align}
where the matrices $Q$ and $U$ are defined according to
\begin{equation}
     Q = \begin{pmatrix} 0 & b_{++} \\ B_{--} & 0 \end{pmatrix} \qd, \qd
     U = \begin{pmatrix} - B_{+-} & b_{++} \\ - B_{--} & B_{+-}
         \end{pmatrix}\ .
\end{equation}

Mutual compatibility of the linear differential equations 
(\ref{lindiff}) leads to a set of nonlinear partial differential
equations for the potentials $b_{++}$ and $B_{--}$ \cite{GIK98}.
In particular, the space and time evolution is driven by the
separated nonlinear Schr\"o\-dinger equation,
\begin{align} \label{sepnls1}
     \i \6_t b_{++} & = - \tst{\2} \6_x^2 b_{++} - 4 b_{++}^2 B_{--}\ ,
                          \\ \label{sepnls2}
     \i \6_t B_{--} & = \tst{\2} \6_x^2 B_{--} + 4 B_{--}^2 b_{++}\ .
\end{align}

\section{Connection between Fredholm determinant and potentials}
\label{sec:taufun}
To describe the correlation functions (\ref{gp2}) and (\ref{gm2})
one has to relate the Fredholm determinant $\det(\hat I + \hat V)$
and the potentials $B_{ab}$ and $C_{ab}$. Let us use the abbreviation
$\s (x,t,\be) = \ln \det(\hat I + \hat V)$. The logarithmic derivatives
of the Fredholm determinant with respect to $x$, $t$ and $\be$ are
\begin{align} \label{logdetx}
     \6_x \s = & - 2\i B_{+-}\ , \\ \label{logdett}
     \6_t \s = & - 2\i (C_{+-} + C_{-+} + G(x,t) B_{--})\ , \\
     \6_\be \s = & - 2\i t \6_\be (C_{+-} + C_{-+} + G(x,t) B_{--})
                     - 2\i x \6_\be B_{+-} - 2(\6_\be B_{+-})^2
		     \nn \\ \label{logdetb}
               &    - 2\i t (B_{--} \6_\be b_{++}
		     - b_{++} \6_\be B_{--})
	             + 2(\6_\be b_{++})(\6_\be B_{--})\ .
\end{align}
For the calculation of the asymptotics of the Fredholm determinant
we further need the second derivatives of $\s$ with respect to space
and time,
\begin{align} \label{logdetxx}
     \6_x^2 \s = & 4 B_{--} b_{++}\ , \\ \label{logdetxt}
     \6_x \6_t \s = & 2\i (B_{--} \6_x b_{++} - b_{++} \6_x B_{--})\ ,
                        \\ \label{logdettt}
     \6_t^2 \s = & 2\i (B_{--} \6_t b_{++} - b_{++} \6_t B_{--})
                     + 8 B_{--}^2 b_{++}^2 + 2 (\6_x B_{--})
		     (\6_x b_{++})\ .
\end{align}
Note that
\begin{equation} \label{limdhb}
     \lim_{\be \rightarrow - \infty} \s = 0 \qd.
\end{equation}
This follows from $\lim_{\be \rightarrow - \infty} \dh (\la) = 0$
and is important for fixing the integration constant in the calculation
of the asymptotics of the determinant.

\section{The Riemann-Hilbert problem} \label{sec:rhp}
From now on we will restrict ourselves to the case of negative
effective chemical potential, $\be < 0$. Recall that this is the
condition for the system to be in the gas phase. For negative $\be$
the logarithmic derivatives $\6_x \s$ and $\6_t \s$ of the Fredholm
determinant and the potentials $b_{++}$ and $B_{--}$ are determined
by the following matrix Riemann-Hilbert problem, which was derived
from the determinant representation (see section \ref{sec:detrep})
in \cite{GIKP98}.
\begin{enumerate}
\item
$\Ph : \mathbb{C} \rightarrow \End(\mathbb{C}^2)$ is analytic in
$\mathbb{C} \setminus \mathbb{R}$.
\item
$\lim_{\la \rightarrow \infty} \Ph(\la) = I_2$.
\item
$\Ph$ has a discontinuity across the real axis described by the
condition
\begin{equation} \label{conmat}
     \Ph_- (\la) = \Ph_+ (\la)
                   \begin{pmatrix}
		      1 & p(\la) e^{- 2 \tau(\la)} \\
		      q(\la) e^{2 \tau(\la)} & 1 + p(\la) q(\la)
		   \end{pmatrix}
\end{equation}
for all $\la \in \mathbb{R}$.
\end{enumerate}
Here $I_2$ denotes the $2 \times 2$ unit matrix. The functions $p(\la)$
and $q(\la)$ are defined as
\begin{align}
     p(\la) = & \i (\cos(\h) - 1)(1 - \dh(\la)) \a_+(\la) \a_-(\la)\ ,\\
     q(\la) = & - \, \frac{2 \i \dh(\la)}{\a_+(\la) \a_-(\la)}\ ,
\end{align}
where
\begin{equation}
     \a(\la) = \exp \left\{ - \, \frac{1}{2 \p \i}
               \int_{- \infty}^\infty \frac{d \m}{\m - \la} \,
	       \ln\big(1 + \dh(\m)(e^{- \i \h} - 1)\big) \right\}\ .
\end{equation}

The functions $\6_x \s$, $\6_t \s$, $b_{++}$ and $B_{--}$ can be
expressed through the coefficients in the asymptotic expansions of
$\Ph(\la)$ and $\ln(\a(\la))$ for large spectral parameter $\la$.
Let
\begin{equation}
     \Ph(\la) = I_2 + \frac{\Ph^{(1)}}{\la}
                    + \frac{\Ph^{(2)}}{\la^2}
		    + \CO \left( \frac{1}{\la^3} \right)
\end{equation}
and
\begin{equation}
     \ln(\a(\la)) = \frac{\a_1}{\la} + \frac{\a_2}{\la^2}
		    + \CO \left( \frac{1}{\la^3} \right)\ .
\end{equation}
Then
\begin{align} \label{dsigmaas}
     \6_x \s = 2 \i \a_1 + \i \; \tr\{\Ph^{(1)} \s^z \} \qd , & \qd
     \6_t \s = 4 \i \a_2 + 2 \i \; \tr\{\Ph^{(2)} \s^z \}\ , \\
     b_{++} = \Ph^{(1)}_{12} \qd, & \qd
     B_{--} = - \Ph^{(1)}_{21}\ .
\end{align}

The Riemann-Hilbert problem is the appropriate starting point for the
asymptotic analysis of the potentials $b_{++}$ and $B_{--}$ which determine
the asymptotics of the two-particle Green functions $G_{\auf \auf}^\pm$.
For impenetrable Bosons a similar analysis was carried out in \cite{IIKV92}.
Fortunately, the result of \cite{IIKV92} depends only on some general
properties of the functions $p(\la)$ and $q(\la)$ entering the conjugation
matrix in (\ref{conmat}), and also applies in the present case.
Alternatively, the non-linear steepest descent method of Deift and Zhou
\cite{DeZh93} could be applied.

\section{Asymptotics of the correlation functions} \label{sec:assanal}
The direct asymptotic analysis of the Riemann-Hilbert problem yields the
leading order asymptotics ($x, t \rightarrow \infty$ for fixed ratio
$\la_0 = - 2x/t$) of the functions $\6_x \s$, $\6_t \s$, $b_{++}$ and
$B_{--}$ \cite{GIK98,IIKV92}. It turns out, in particular, that $b_{++}$ and
$B_{--}$ are a decaying solution of the separated nonlinear Schr\"odinger
equation (\ref{sepnls1}), (\ref{sepnls2}). Now the form of the complete
asymptotic decomposition of the decaying solutions of the separated
nonlinear Schr\"odinger equation is known \cite{SeAb76,AbSeBo}.
\begin{align} \label{bppexp}
     b_{++} = & t^{- \2} \left( u_0 + \sum_{n=1}^\infty
                  \sum_{k=0}^{2n} \frac{\ln^k 4t}{t^n} \, u_{nk} \right)
		  \exp \left\{ \frac{\i x^2}{2t} - \i \n \ln 4t \right\}\ ,
		  \\ \label{bmmexp}
     B_{--} = & t^{- \2} \left( v_0 + \sum_{n=1}^\infty
                  \sum_{k=0}^{2n} \frac{\ln^k 4t}{t^n} \, v_{nk} \right)
		  \exp \left\{ - \frac{\i x^2}{2t} + \i \n \ln 4t
		  \right\}\ ,
\end{align}
where $u_0$, $v_0$, $u_{nk}$, $v_{nk}$ and $\n$ are functions of
$\la_0 = -x/2t$ and of $\be$ and $\h$. Inserting the asymptotic
expansions for $B_{--}$ and $b_{++}$ into the differential equations
(\ref{sepnls1}), (\ref{sepnls2}) we obtain expressions for $u_{nk}$,
$v_{nk}$ and $\n$ in terms of $u_0$ and $v_0$, i.e.\ the two unknown
functions $u_0$ and $v_0$ determine the whole asymptotic expansion
(\ref{bppexp}), (\ref{bmmexp}). But $u_0$ and $v_0$ are obtained from
the asymptotic analysis of the Riemann-Hilbert problem (for the
explicit expressions see \cite{GIK98}). Hence we know, in principle, the
complete asymptotic decomposition of the potentials $b_{++}$ and $B_{--}$.

In order to obtain the asymptotics of the two-particle Green functions
we still need the asymptotics of the Fredholm determinant. The Fredholm
determinant is related to $b_{++}$ and $B_{--}$ through equations
(\ref{logdetxx})-(\ref{logdettt}) and (\ref{logdetx})-(\ref{logdetb}).
We may integrate (\ref{logdetxx})-(\ref{logdettt}) to obtain the
asymptotic expansions of $\6_x \s$ and $\6_t \s$. The integration
constant is a function of $\be$. It is fixed by the leading asymptotics,
which, using (\ref{dsigmaas}), can be obtained from the direct
asymptotic analysis of the Riemann-Hilbert problem. Then, integrating
(\ref{logdetx})-(\ref{logdetb}) yields $\s$ up to a numerical constant,
which follows from the asymptotic condition (\ref{limdhb}). The
calculation is the same as for the impenetrable Bose gas and can be
found on pages 455-457 of \cite{KBIBo}.

Finally, we obtain the following expressions for the leading asymptotics
of the correlation function,
\begin{align} \label{gp} \nn
     g^+ (x,t) = & 
	e^{\i x^2/2t + 2 \i t \be} e^{- 2 \i t (h + \ln(2\ch (h)))}
        \int_{- \p}^\p \! d\h \, \frac{F(\g,\h)}{1 - \cos(\h)} \cdot
        \\ & \cdot
	C^+ (\la_0,\be,\h) \, (4t)^{\2 (\n - \i)^2} \,
        \exp \left\{ \frac{1}{\p} \int_{-\infty}^\infty \! d \la \,
             |x + 2 \la t| \ln(\ph(\la,\be)) \right\}\ , \\[2ex]
     \label{gm} \nn
     g^- (x,t) = &
	e^{- \i x^2/2t - 2 \i t \be} e^{2 \i t (h + \ln(2\ch (h)))}
        \int_{- \p}^\p \! d\h \, \frac{F(\g,\h)}{2 \g} \cdot
	\\ & \cdot
        C^- (\la_0,\be,\h) \, (4t)^{\2 (\n + \i)^2} \, 
        \exp \left\{ \frac{1}{\p} \int_{-\infty}^\infty \! d \la \,
             |x + 2 \la t| \ln(\ph(\la,\be)) \right\}\ ,
\end{align}
where
\begin{align}
     \ph(\la,\be) = & 1 + \dh(\la)\left(
                      e^{- \i \h \sign(\la - \la_0)} - 1 \right)\ , \\
     \n = & - \frac{1}{2 \p} \ln \left( 1 - 2 (1 - \cos(\h)) \dh (\la_0)
              (1 - \dh (\la_0)) \right)\ , \\ \nn
     C^+ (\la_0,\be,\h) = & - \, |\sin(\h/2)| \frac{\sqrt{\n}}{2 \p}
        \exp \left\{ \2 (\la_0^2 - \be) + \i \Ps_0 + \frac{\n^2}{2}
	\right. \\ & \qd
	- \int_{- \infty}^\be \! d \be \, (\i \nu /2 + \n \6_\be \Ps_0)
	\\ \nn & \qd
        \left. + \frac{1}{2 \p^2} \int_{- \infty}^\be \! d \be \,
	         \left( \6_\be \int_{- \infty}^\infty \! d \la \,
		 \sign(\la - \la_0) \ln( \ph(\la,\be)) \right)^2
		 \right\}\ , \\[1ex] \label{cmm}
     C^- (\la_0,\be,\h) = & C^+ (\la_0,\be,\h) \,
                              \exp(- (\la_0^2 - \be) - 2 \i \Ps_0)/
			      \sin^2 (\h/2)\ .
\end{align}
$\la_0 = - x/2t$ is the stationary point of the phase $\tau(\la)$
(i.e.\ $\tau'(\la_0) = 0$), and the functions $\Ps_0$ and $\Ps_1$
are defined as
\begin{align} \label{psinull}
     \Ps_0  = & - \frac{3 \p}{4} + \mbox{arg} \, \G(\i \nu) + \Ps_1\ ,
                  \\ \label{psione}
     \Ps_1  = & - \frac{1}{\p} \int_{- \infty}^\infty \!
	         d \la \; \sign(\la - \la_0) \ln |\la - \la_0|
	         \6_\la \ln(\ph(\la,\be))\ .
\end{align}

Equations (\ref{gp}) and (\ref{gm}) are valid for large $t$ and fixed
finite ratio $\la_0 = - x/2t$. Correlations in the pure space direction
$t = 0$ were discussed in \cite{Berkovich91}. We would like to emphasize
that (\ref{gp}) and (\ref{gm}) still hold for arbitrary temperatures. The
low temperature limit will be discussed in the next section. Note
that there is no pole of the integrand at $\h = 0$, since $\sqrt{\n}
\sim |\h|$ for small $\h$ and thus $C^+ (\la_0,\be,\h) \sim \h^2$.
\section{Asymptotics in the low temperature limit} \label{sec:steepest}
For the following steepest descent calculation we transform the
$\h$-integrals in (\ref{gp}), (\ref{gm}) into complex contour integrals
over the the unit circle, setting $z = e^{\i \h}$. Since we would like
to consider low temperatures, we have to restore the explicit
temperature dependence by scaling back to the physical space and time
variables $x$ and $t$ and to the physical correlation functions
$G_{\auf \auf}^\pm$. Recall that in the previous sections we have
suppressed an index `$r$' referring to `rescaled'. Let us restore this
index in order to define $k_0 = \la_0 \sqrt{T} = x/2t$, $\dh (k) =
\dh_r (k/\sqrt{T})$, $\ph (k,\be) = \ph_r (k/\sqrt{T},\be)$,
$C^\pm (k_0, \be, z) = C^\pm_r (\la_0, \be, \h)$, $F(\g, z) =
F_r (\g, \h)$. Then
\begin{align} \label{gpc}
     G_{\auf \auf}^+ (x,t) = & 2 \i \sqrt{T} \,
        e^{\i x^2/4t + \i t (\m - B)}
	\oint dz \, \frac{F(\g,z)}{(z - 1)^2} \, C^+ (k_0,\be,z)
	   (2Tt)^{\2 (\n (z) - \i)^2} e^{t S(z)}\ , \\ \label{gmc}
     G_{\auf \auf}^- (x,t) = & - \i \sqrt{T} \,
        e^{- \i x^2/4t - \i t (\m - B)}
	\oint dz \, \frac{F(\g,z)}{2 \g z} \, C^- (k_0,\be,z)
	   (2Tt)^{\2 (\n (z) + \i)^2} e^{t S(z)}\ ,
\end{align}
where
\begin{equation}
     S(z) = \frac{1}{\p} \int_{- \infty}^\infty \! dk \, |k - k_0|
            \ln(\ph(k,\be))\ .
\end{equation}

We would like to calculate the contour integrals (\ref{gpc}),
(\ref{gmc}) by the method of steepest descent. For this purpose we
have to consider the analytic properties of the integrands. Let us
assume that $k_0 \ge 0$, and let us cut the complex plane along the
real axis from $- \infty$ to $- e^{- \be}$ and from $- e^{\be -
k_0^2 /T}$ to $0$. The integrands in (\ref{gpc}) and (\ref{gmc})
can be analytically continued as functions of $z$ into the cut plane
with the only exception of the two simple poles of $F(\g,z)$ at
$z = \g^{\pm 1}$. We may therefore deform the contour of integration
as long as we never cross the cuts and take into account the pole
contributions, if we cross $z = \g$ or $z = \g^{-1}$.

The saddle point equation $\6 S/\6 z = 0$ can be represented in the
form
\begin{equation} \label{saddle}
     \int_0^\infty \frac{dk \, k}{1 + z^{-1} e^{- \be}
        e^{(k - k_0)^2 /T}} =
     \int_0^\infty \frac{dk \, k}{1 + z e^{- \be}
        e^{(k + k_0)^2 /T}}\ .
\end{equation}
This equation was discussed in the appendix of \cite{GIKP98}. In
\cite{GIKP98} it was shown that (\ref{saddle}) has exactly one real
positive solution which is located in the interval $[0,1]$. It was
argued that this solution gives the leading saddle point contribution
to (\ref{gpc}) and (\ref{gmc}). At small temperatures (\ref{saddle})
can be solved explicitly. There are two solutions $z_\pm = \pm z_c$,
where
\begin{equation} \label{zc}
     z_c = \frac{T^{3/4}}{2 \p^{1/4} k_0^{3/2}} \, e^{- k_0^2/2T}\ .
\end{equation}
In the derivation of (\ref{zc}) we assumed that $k_0 \ne 0$. The case
$k_0 = 0$ has to be treated separately (see below).

The phase $t S(z)$ has the low temperature approximation
\begin{equation} \label{lowphase}
     t S(z) = - 2 k_0 D t \left\{ \left( 1 - \frac{1}{z} \right)
	          z_c^2 + 1 - z \right\}\ .
\end{equation}
Here $D = \6 P/\6 \m$ is the density of the electron gas. The low
temperature expansion (\ref{lowphase}) is valid in an annulus
$e^{\be - k_0^2/T} \ll |z| \ll e^{- \be}$, which lies in our cut plane.
The unit circle and the circle $|z| = z_c$ are inside this annulus.
We may thus first apply (\ref{lowphase}) and then deform the contour of
integration from the unit circle to the small circle $|z| = z_c$. Let
us parameterize the small circle as $z = z_c e^{\i \a}$, $\a \in
[- \p, \p]$. Then $S(z(\a)) = - 2k_0 D ((z_c - 1)^2 + 2z_c
(1 - \cos(\a)))$, which implies that the small circle is a steepest
descent contour and that on this contour $S(z_-) \le S(z) \le S(z_+)$.
The maximum of $S(z)$ on the steepest descent contour at $z = z_+$ is
unique and therefore provides the leading saddle point contribution to
(\ref{gpc}), (\ref{gmc}) as $t \rightarrow \infty$. The saddle point
approximation becomes good when $t S(z(\a)) = - 2k_0Dt((z_c - 1)^2 +
z_c \a^2 + \CO(\a^4))$ becomes sharply peaked around $\a = 0$. Hence,
the relevant parameter for the calculation of the asymptotics of
$G_{\auf \auf}^\pm$ is $2k_0Dt = xD$ rather than $t$. $xD$ has to be
large compared to $z_c^{-1}$. The parameter $xD$ has a simple
interpretation. It is the average number of particles in the interval
$[0,x]$. Let us consider two different limiting cases.
\begin{enumerate}
\item
$xD \rightarrow 0$, the number of electrons in the interval $[0,x]$
vanishes. In this regime the interaction of the electrons is
negligible. An electron propagates freely from 0 to $x$. $G_{\auf
\auf}^\pm$ cannot be calculated by the method of steepest descent. We
have to use the integral representation (\ref{gp}), (\ref{gm})
instead. Since $t S(z)$ and $\nu(z)$ tend to zero on the contour of
integration, the integrals in (\ref{gp}) and (\ref{gm}) are easily
calculated. We find $G_{\auf \auf}^\pm = G_f^\pm$ (see (\ref{gpf}),
(\ref{gmf})), which is the well known result for free Fermions.
\item
$xD \gg z_c^{-1}$, the average number of electrons in the interval
$[0,x]$ is large. This is the {\it true asymptotic region},
$x \rightarrow \infty$. In this region the interaction becomes
important. At the same time the method of steepest descent can be used
to calculate $G_{\auf \auf}^\pm$. This case will be studied below.
\end{enumerate}

In the process of deformation of the contour from the unit circle to
the small circle of radius $z_c$ we may cross the pole of the function
$F(\g,z)$ at $z = \g^{-1}$. Then we obtain a contribution of the pole
to the asymptotics of $G_{\auf \auf}^\pm$.  It turns out that the pole
contributes to $G_{\auf \auf}^\pm$, when the magnetic field is below a
critical positive value, $B_c = k_0^2/4$. Below this value the pole
contribution always dominates the contribution of the saddle point.
Hence, we have to distinguish two different asymptotic regions,
$B > B_c$ and $B < B_c$. On the other hand, if we consider the
asymptotics for fixed magnetic field, we have to treat the cases
$B > 0$ and $B \le 0$ separately. For $B > 0$ we have to distinguish
between a time like regime ($k_0^2 < 4B$) and a space like regime
($k_0^2 > 4B$). In these respective regimes we obtain the asymptotics
(\ref{gpmas2}), (\ref{gpmas}).

In the limit $B \rightarrow - \infty$, $\m \rightarrow - \infty$,
$\m - B$ fixed there are no $\ab$-spin electrons left in the system,
$D_\ab \rightarrow 0$. This is the free Fermion limit. In the free
Fermion limit $B < B_c$, and the asymptotics of $G_{\auf \auf}^+ (x,t)$
and $G_{\auf \auf}^- (x,t)$ are given by the equations (\ref{gpmas}),
which turn into the expressions (\ref{gpf}), (\ref{gmf}) for free
Fermions.

The pure time direction $k_0 = 0$ requires a separate calculation.
For $k_0 = 0$ the saddle point equation (\ref{saddle}) has the solutions
$z = \pm 1$ for all temperatures. The unit circle is a steepest descent
contour with unique maximum of $S(z)$ at $z = 1$, which gives the
leading asymptotic contribution to the integrals in (\ref{gpc}) and
(\ref{gmc}). We find algebraically decaying correlations,
\begin{equation}
     G_{\auf \auf}^+ (0,t) = C_0^+ t^{-1} e^{\i t (\m - B)} \qd, \qd
     G_{\auf \auf}^- (0,t) = C_0^- t^{-1} e^{- \i t (\m - B)}\ ,
\end{equation}
where
\begin{align}
     C_0^+ = & \frac{e^{- \i \frac{\p}{4}}}
                 {2 \sqrt{2 \p T}}
                 (1 + 2 e^{- 2B/T}) \nn \\ &
		 \left[ (e^{(\m + B)/T} + e^{(\m - B)/T})
		 (1 + e^{(\m + B)/T} + e^{(\m - B)/T}) \right]^{- \2}\ , \\
     C_0^- = & \frac{e^{\i \frac{\p}{4}}}
                 {2 \sqrt{2 \p T}} \,
                 \frac{1 + 2 e^{- 2B/T}}{1 + e^{2B/T}}
		 \left[ \frac{e^{(\m + B)/T} + e^{(\m - B)/T}}{
		 1 + e^{(\m + B)/T} + e^{(\m - B)/T}} \right]^\2\ .
\end{align}
These formulae are valid at any temperature.
\subsection*{Acknowledgment}
The results presented in this talk were obtained in collaboration with
A. R. Its and V. E. Korepin to whom the author would like to express his deep
gratitude. The author thanks W. Pesch for a critical reading of the
manuscript. He was supported by H. Fehske and the DFG Schwerpunktprogramm
SPP 1073 `Kollektive Quantenzust\"ande in elektronisch eindimensionalen
\"Ubergangsmetallverbindungen'.


\begin{thebibliography}{10}

\bibitem{Yang67}
C.~N. Yang, Phys. Rev. Lett. {\bf 19}, 1312 (1967).

\bibitem{Gaudin67}
M.~Gaudin, Phys. Lett. A {\bf 24}, 55 (1967).

\bibitem{GoKo99a}
F.~G\"ohmann and V.~E. Korepin, Phys. Lett. A {\bf 260}, 516 (1999).

\bibitem{GIKP98}
F.~G\"ohmann, A.~G. Izergin, V.~E. Korepin, and A.~G. Pronko, Int. J. Mod.
  Phys. B {\bf 12}, 2409 (1998).

\bibitem{GIK98}
F.~G\"ohmann, A.~R. Its, and V.~E. Korepin, Phys. Lett. A {\bf 249}, 117
  (1998).

\bibitem{Haldane80}
F.~D.~M. Haldane, Phys. Rev. Lett. {\bf 45}, 1358 (1980).

\bibitem{Haldane81}
F.~D.~M. Haldane, J. Phys. C {\bf 14}, 2585 (1981).

\bibitem{Voit94}
J.~Voit, Rep.\ Prog.\ Phys. {\bf 57}, 977 (1994).

\bibitem{BPZ84}
A.~A. Belavin, A.~M. Polyakov, and A.~B. Zamolodchikov, Nucl. Phys. B {\bf
  241}, 333 (1984).

\bibitem{LiWu68}
E.~H. Lieb and F.~Y. Wu, Phys. Rev. Lett. {\bf 20}, 1445 (1968).

\bibitem{FrKo90}
H.~Frahm and V.~E. Korepin, Phys. Rev. B {\bf 42}, 10553 (1990).

\bibitem{FrKo91}
H.~Frahm and V.~E. Korepin, Phys. Rev. B {\bf 43}, 5653 (1991).

\bibitem{DEGKKK00}
T.~Deguchi, F.~H.~L. Essler, F.~G\"ohmann, A.~Kl\"umper, V.~E. Korepin, and
  K.~Kusakabe, Phys. Rep. {\bf 331}, 197 (2000).

\bibitem{Takahashi72}
M.~Takahashi, Prog. Theor. Phys. {\bf 47}, 69 (1972).

\bibitem{Takahashi74}
M.~Takahashi, Prog. Theor. Phys. {\bf 52}, 103 (1974).

\bibitem{Takahashi71b}
M.~Takahashi, Prog. Theor. Phys. {\bf 46}, 1388 (1971).

\bibitem{IzPr97}
A.~G. Izergin and A.~G. Pronko, Phys. Lett. A {\bf 236}, 445 (1997).

\bibitem{IzPr98}
A.~G. Izergin and A.~G. Pronko, Nucl. Phys. B {\bf 520}, 594 (1998).

\bibitem{IIKS90}
A.~R. Its, A.~G. Izergin, V.~E. Korepin, and N.~Slavnov, Int. J. Mod. Phys. B
  {\bf 4}, 1003 (1990).

\bibitem{KBIBo}
V.~E. Korepin, N.~M. Bogoliubov, and A.~G. Izergin, {\em Quantum Inverse
  Scattering Method and Correlation Functions}, Cambridge University Press,
  (1993).

\bibitem{IIKV92}
A.~R. Its, A.~G. Izergin, V.~E. Korepin, and G.~G. Varzugin, Physica D {\bf
  54}, 351 (1992).

\bibitem{DeZh93}
P.~A. Deift and X.~Zhou, Ann. of Math. {\bf 137}, 295 (1993).

\bibitem{SeAb76}
H.~Segur and M.~J. Ablowitz, J. Math. Phys. {\bf 17}, 710 (1976).

\bibitem{AbSeBo}
M.~J. Ablowitz and H.~Segur, {\em Solitons and the Inverse Scattering
  Transform}, SIAM, Philadelphia,  (1981).

\bibitem{Berkovich91}
A.~Berkovich, J. Phys. A {\bf 24}, 1543 (1991).

\end{thebibliography}

\end{document}